\documentclass[prl,twocolumn,showpacs,amsmath,amssymb]{revtex4-1}
\usepackage{graphicx}       
\usepackage{mathtools}
\usepackage{epsfig}
\usepackage{xcolor}
\usepackage{dcolumn}        
\usepackage{ulem}
\newcommand{\qvec}{{\bf q}}
\newcommand{\beq}{\begin{equation}}
\newcommand{\eeq}{\end{equation}}
\newcommand{\beqa}{\begin{eqnarray}}
\newcommand{\eeqa}{\end{eqnarray}}

\begin{document}
\title{Dissipation-driven strange metal behavior}
\author{S. Caprara$^{1,2}$, C. Di Castro$^1$, G. Mirarchi$^1$, G. Seibold$^3$, M. Grilli$^{1,2}$}
\affiliation{$^1$Dipartimento di Fisica, Universit\`a di 
Roma ``La Sapienza'', P.$^{le}$ Aldo Moro 5, 00185 Roma, Italy}

\affiliation{$^2$ISC-CNR, Unit\`a di Roma ``Sapienza''}

\affiliation{$^3$ Institut f{\"u}r Physik, BTU Cottbus-Senftenberg - PBox 101344, D-03013 Cottbus, Germany} 

\begin{abstract}
{Anomalous metallic properties are often observed in the proximity of quantum critical points (QCPs), with violation 
of the Fermi Liquid paradigm. We propose a scenario where, due to the presence of a nearby QCP, dynamical fluctuations 
of the order parameter with {\it finite correlation length} mediate a nearly isotropic scattering among the  
quasiparticles over the entire Fermi surface. This scattering produces an anomalous metallic behavior, which is
extended to the lowest temperatures by an increase of the damping of the fluctuations. We phenomenologically identify 
one single parameter ruling this increasing damping when the temperature decreases, accounting for both the 
linear-in-temperature resistivity and the seemingly divergent specific heat observed, e.g., in high-temperature 
superconducting cuprates and some heavy-fermion metals.
}
\end{abstract}
\date{\today}
\maketitle

{\it --- Introduction ---}
Landau's Fermi Liquid (FL) theory is one of the most successful paradigms in condensed matter physics and usually 
describes very well the prominent properties of metals even in cases when the interaction is strong, like, e.g., 
in heavy-fermions metals or in the normal (i.e., non superfluid) phase of $^3$He. However, in the last decades, a 
wealth of systems violating the paradigmatic behavior has been discovered. In particular, it has been noticed that 
in several different materials, like heavy-fermions metals \cite{stewart} and iron-based superconductors \cite{Walmsley}, 
a non-FL behavior can occur in the proximity of quantum critical points (QCPs), i.e., near zero-temperature second-order 
phase transitions, where the uniform metallic state is unstable towards some ordered state. It is worth mentioning
that, apart from the paradigmatic case of the one dimensional Luttinger liquid, there are also theories for the violation 
of the FL behavior that do not rely on an underlying criticality \cite{anderson,kastrinakis}. In some cases, like in 
high-temperature superconducting cuprates (henceforth, cuprates), the ordered state may be unaccomplished due to disorder, 
low dimensionality, and/or competition with other phases, like superconductivity. Nevertheless, the non-FL behavior 
is observed also in these cases of {\it missed} quantum criticality, showing that a mere tendency to order and the 
presence of abundant order parameter fluctuations (henceforth, fluctuations) may be sufficient to create a non-FL state. 
The general underlying idea is that the fluctuations are intrinsically dynamical, with a characteristic energy $m$ 
becoming smaller and smaller as the correlation length $\xi$ grows larger and larger, when the QCP is approached. In 
the paradigmatic case of a gaussian QCP in a metal, with a dynamical critical index $z=2$, the propagator of 
the fluctuations with wavevector $\qvec$ and frequency $\omega$ is \cite{CDG-ZP}
\beq
D(\qvec,\omega)=\left(m+\bar\nu |\qvec -\qvec_c|^2-\omega^2/\overline{\Omega}-\mathrm i\gamma \omega\right)^{-1},
\label{fluctuator}
\eeq
where $m=\bar\nu \xi^{-2}$ is the {\it mass} of the fluctuations, $\bar\nu$ is typically an electron energy scale [we 
work with dimensionless momenta, measured in reciprocal lattice units (r.l.u.) $2\pi/a$], $\qvec_c$ is the critical 
wave vector, and $\overline{\Omega}$ is a frequency cutoff. A crucial role in the following will be played by the 
imaginary term in the denominator, which describes the Landau damping of the fluctuations, as they decay in particle-hole 
pairs. The dimensionless parameter $\gamma$ is usually proportional to the electron density of states, which sets a 
measure of the phase space available for the decay of the fluctuations. Clearly, in the gaussian case, for $\omega=0$ 
and $\qvec\approx\qvec_c$ one obtains the standard Ornstein-Zernike form of the static susceptibility. 
The same behavior of the fluctuations can be obtained within a time-dependent Landau-Ginzburg approach, 
where $\gamma$ is the coefficient of the time derivative and the decay rate of the fluctuations is given by 
$\tau_\qvec^{-1}=(m+\bar\nu |\qvec -\qvec_c|^2)/\gamma$.

Approaching the QCP, $\xi$ grows, $m$ decreases and the fluctuations become softer and softer, thereby mediating a stronger 
and stronger interaction between the fermion quasiparticles (henceforth, simply quasiparticles). In two and three dimensions 
the interaction {\it could} be strong enough to destroy the FL state \cite{metzner-1990}. For ordering at finite wavevectors, 
though, there is a pitfall in this scheme \cite{hlubina-1990}: due to momentum conservation, this singular low-energy 
scattering only occurs between quasiparticles near points of the Fermi surface that are connected by $\qvec_c$ (hot spots). 
All other regions are essentially unaffected by this singular scattering and most of the quasiparticles keep their standard 
FL properties. As a result, for instance in transport, a standard FL behavior would occur, with a $T^2$ FL-like resistivity 
\cite{hlubina-1990}. Disorder may help to blur and enlarge the hot regions \cite{rosch-1990}, but it does not completely 
solve the above difficulty. Of course, this limitation does not occur in cases where $\qvec_c={\bf 0}$ (like, e.g, near a 
ferromagnetic \cite{belitz-kirkpatrick}, or a circulating-current \cite{varma}, or a nematic \cite{kivelson-scheurer} QCP), 
or near a local QCP (i.e., when the singular behavior persists locally for all $\qvec$) \cite{qimiaosi,coleman,burdin,sachdev-syk}. 
However, the very fact that similar non-FL behavior also occurs near QCPs with finite $\qvec_c$ calls for a revision of the 
above scheme searching for a general and robust way to account for non-FL phases irrespective of the ordering wave vector. 
 
The main goal of the present work is to describe an alternative scenario for the non-FL behavior, based on the idea 
that the decay rate of the fluctuations  $\tau_\qvec^{-1}$ becomes very small not only at $\qvec \approx \qvec_c$, 
because of a diverging $\xi$, but rather at all $\qvec$'s, because of a (nearly) diverging $\gamma$, as $T$ goes to 
zero at special values of the control parameter as, e.g., doping in cuprates. We will adopt a phenomenological approach 
and will not exhibit a microscopic mechanism inducing this growth of dissipation. However, we will explicitly show that 
a finite $\xi$ and a large $\gamma$ are generic sufficient conditions to obtain the most prominent signatures of 
non-FL behavior: a linear-in-temperature ($T$) resistivity (even down to very low temperature) and a (seemingly) diverging 
specific heat. For the sake of concreteness, we will consider the paradigmatic case of cuprates, where at some specific 
doping both features are observed \cite{legros-2019,michon-2019}, having in mind that they also commonly occur in many 
other systems like, e.g. heavy fermions \cite{stewart}. This suggests that our proposal might have a broad applicability.
 
{\it --- Dissipation-driven strange metal behavior ---}
The above scenario can be achieved on the basis of three simple and related ingredients: (a) The proximity to a QCP, 
bringing the fluctuations to sufficiently low energy; (b) Some quenching mechanism preventing the full development 
of criticality so that the mass $m$ and the other parameters of the dynamical fluctuations do not vary in a significant 
way with temperature; (c) Some mechanism driving an increase of the Landau damping parameter $\gamma$. Indeed the 
non-FL behavior persists down to a temperature scale 
$T_{FL}\sim \omega_0\equiv m/\gamma =\bar{\nu}/(\xi^2 \gamma)=\tau_{\qvec_c}^{-1}$ when $\xi$ is {\it finite and 
not particularly large}. In cuprates, recent resonant X-ray scattering (RXS) experiments \cite{arpaia-2019} show 
that conditions (a) and (b) hold: the occurrence of a temperature dependent narrow peak due to charge density waves 
testifies the proximity to a QCP (although hidden and not fully attained due to the competition with the superconducting 
phase). The concomitant occurrence of a broad peak witnesses for the presence of dynamical charge density fluctuations 
with rather short correlation length and broad momentum distribution. These abundant charge density fluctuations are 
available to isotropically scatter the quasiparticles over a broad range of momenta and no clear distinction can be 
done between hot and cold Fermi surface regions \cite{seibold-2020}. This was the first explicit example that a 
quenched criticality with a finite ordering wave vector $\qvec_c$ can still give rise to strong but isotropic 
scattering, thereby bypassing the problem that in standard hot spot models most electrons contribute with a $\sim T^2$ 
scattering rate to transport \cite{hlubina-1990}. This shows that conditions (a) and (b) are enough to account for a 
linear-in-$T$ resistivity above $T_{FL}$. Condition (c) becomes instead mandatory because an increasing $\gamma$ is 
needed to extend to lower temperatures the non-FL behavior, accounting for the persistence of the linear resistivity 
observed down to a few kelvins, which is the so-called Planckian behavior \cite{legros-2019}, as well as a seemingly 
diverging specific heat \cite{michon-2019} (see below).  

\begin{figure}
\includegraphics[angle=0,scale=0.45]{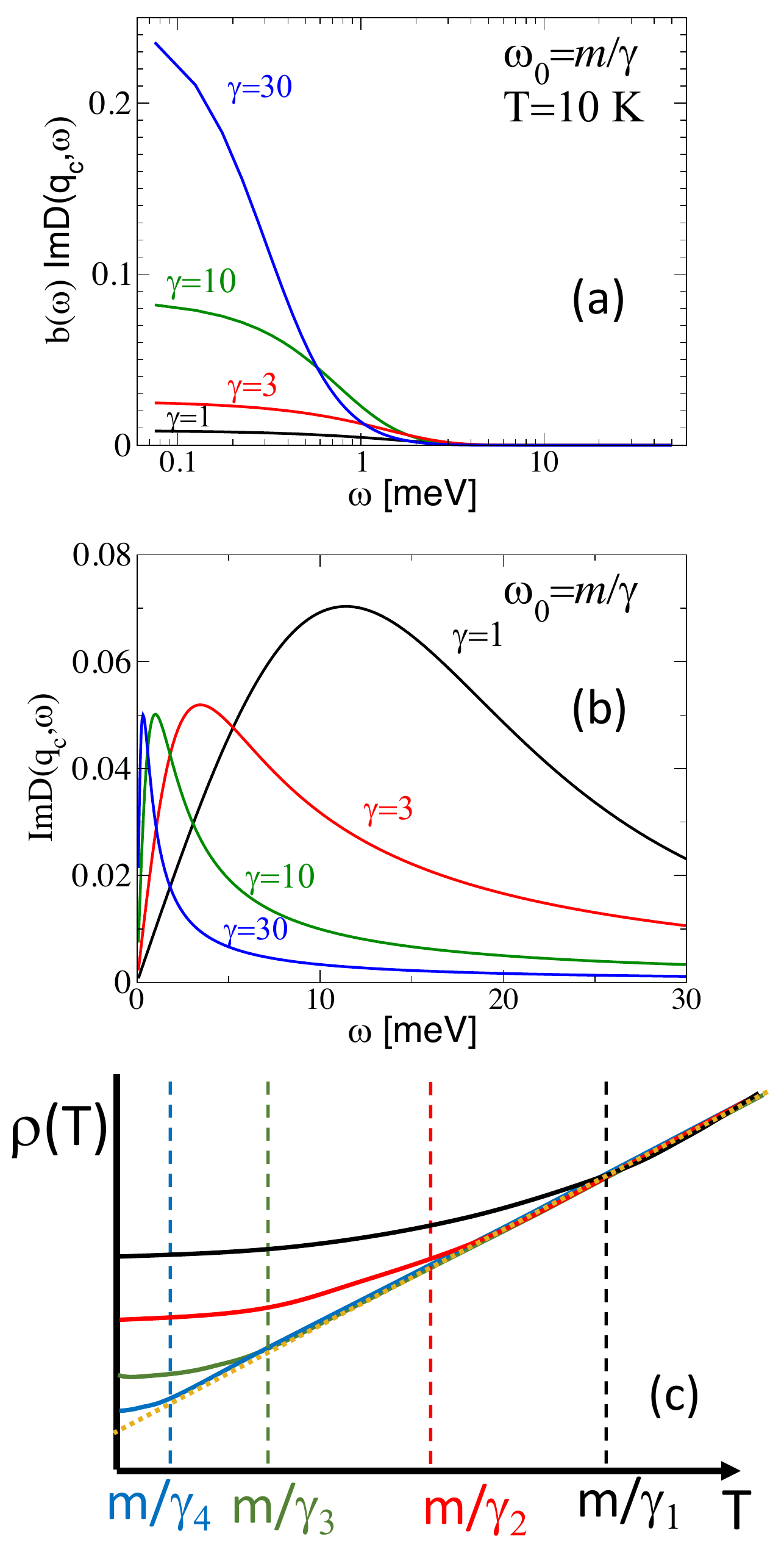}
\caption{Sketch of the shift induced by an increasing damping parameter $\gamma=1\to 30$ on the fluctuation spectral 
function $\mathrm{Im}\, D$ in the presence (a)  and in the absence (b) of a Bose thermal distribution; (c) sketch 
of the effect on the resistivity induced by the decrease of the characteristic energy $\omega_0\equiv m/\gamma$ of 
the fluctuations responsible for the quasiparticle scattering, with the scattering rate given by the imaginary part 
of the electron self-energy, computed at second order in the coupling $g$ between electron quasiparticles and fluctuations. 
The orange dotted line represents the Planckian limit (linearity down to $T=0$), corresponding to a 
divergent $\gamma$.}
\label{fig-0}
\end{figure}
 
To address this issue, we investigate the effects of an increasing $\gamma$ on the fluctuations, that provide 
both a scattering mechanism for resistivity and low-energy excitations for the specific heat. We consider the 
spectral density of the fluctuations \cite{CDG-1995,andergassen-2001,caprara-2002,enss-2007}
\[
\mathrm{Im}\,D(\qvec,\omega)=\frac{\gamma\omega}{\left(m+\bar\nu |\qvec -\qvec_c|^2-\omega^2/\overline{\Omega}\right)^2
+\gamma^2 \omega^2},
\]
which, for $\qvec=\qvec_c$, is maximum at $\omega\approx\omega_0\equiv m/\gamma$. For 
large $\gamma$ (whatever the reason), $\omega_0$ is much smaller than $m$ and sets the 
characteristic energy scale of the dynamical fluctuations. As mentioned above, a large $\gamma$ 
suppresses the energy scales associated with $\tau_\qvec^{-1}$ at all $\qvec$'s.

Fig.\,\ref{fig-0} (a) and (b) display this shift to lower frequencies of $b(\omega)\,\mathrm{Im}\, D(\omega)$ 
and $\mathrm{Im}\, D(\omega)$ when $\gamma$ increases [$b(\omega)=(\mathrm e^{\omega/T}-1)^{-1}$ being the Bose 
function]. Panel (c) schematically shows the corresponding extension of the linear resistivity down to lower and 
lower temperatures. Indeed, although the collective fluctuations obey the Bose statistics, at any temperature 
$T>\omega_0$ they acquire a semiclassical character and their thermal Bose distribution becomes linear in $T$, 
$b(\omega)\approx T/\omega$. Notice that this is the usual situation for phonons when $T$ is above their Debye 
temperature. The only difference here is that a small/moderate $m$ (due to the proximity 
to a QCP) and the large $\gamma$ conspire to render the Debye scale of the fluctuations particularly 
small or even vanishing if $\gamma$ may diverge, while $m$ stays finite. Notice also that the integrated weight of 
the thermally excited fluctuations, $\int d\omega \, b(\omega)\,\mathrm{Im}\, D(\omega)$, depends only very weakly 
on $\gamma$.

\begin{figure}
\includegraphics[angle=0,scale=0.3]{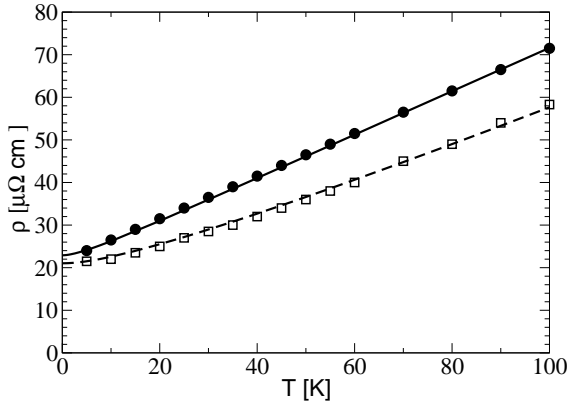}
\caption{Resistivity calculations for a Nd-La$_{2-x}$Sr$_x$CuO$_4$ sample with   $x=0.24$ (black solid line) and for a
 Eu-La$_{2-x}$Sr$_x$CuO$_4$ sample with  $0.24$ (black dashed line). The symbols refer to the corresponding 
 experimental data extracted from Ref.\,\onlinecite{michon-2019}: black filled circles for Nd-La$_{2-x}$Sr$_x$CuO$_4$ 
and  empty black squares for Eu-La$_{2-x}$Sr$_x$CuO$_4$. For the fitting we used a quasiparticle-charge fluctuations coupling 
and the elastic scattering rate due to quenched disorder  
$g^2=0.0351$ and $\Gamma_0=13.4$ meV for Nd-La$_{2-x}$Sr$_x$CuO$_4$ and $g^2=0.0398$ and $\Gamma_0=12.3$ meV for Eu-La$_{2-x}$Sr$_x$CuO$_4$.
}
\label{fig-1}
\end{figure}

{\it --- Resistivity in cuprates ---} In Fig.\,\ref{fig-1} we report our calculation and the experimental data for 
Nd-La$_{2-x}$Sr$_x$CuO$_4$ samples with $x=0.22,0.24$ (data extracted from Ref.\,\onlinecite{michon-2019}). We 
calculate the resistivity by solving the Boltzmann equation, where the scattering rate was obtained from the imaginary 
part of the electron self-energy, computed at second order in the coupling $g$ between electron quasiparticles and 
charge density fluctuations of the form given by Eq.\,(\ref{fluctuator}).

This calculation follows closely the approach  used in Ref.\,\onlinecite{seibold-2020}
for the fermionic tight-binding dispersion, the calculation of the electron scattering rate, 
and the solution of the Boltzmann equation.
Regarding the parameters of the fluctuations, these
were extracted from RXS experiments on a NdBa$_2$Cu$_3$O$_{7-y}$ sample, consistently leading 
to a deviation from linearity below $T_{FL}\approx 100$\,K in agreement with the 
resistivity data. Here, we consider the case of Nd-La$_{2-x}$Sr$_x$CuO$_4$, where resistivity under strong magnetic 
fields is linear down to $T\approx 5$\,K. Unfortunately, although RXS experiments recently confirmed also for these 
cuprates the presence of charge density fluctuations with broad momentum distribution \cite{miao-2020}, detailed data 
are not available to extract their parameters. This is why we assume here that the parameters fitted from RXS data 
in NdBa$_2$Cu$_3$O$_{7-y}$ are still reasonable estimates for Nd-La$_{2-x}$Sr$_x$CuO$_4$ and we therefore 
use similar values: $m=10$\,meV, $\bar\nu=10^3$ meV, $\bar\Omega=30$ meV. 
These values correspond to a rather short coherence length of few lattice spacings  $\xi^{-1}=\sqrt{m/\bar\nu}\approx 0.1$\,r.l.u.). 
We reiterate here that such a short coherence length of the charge density fluctuations is a crucial feature to obtain a nearly 
isotropic scattering over the Fermi surface, so that all quasiparticles are nearly equally scattered and their FL properties 
are uniformly spoiled. As far as the dissipation parameter is concerned, 
we adopt here a phenomenological form for the damping parameter
\beq
\gamma(p,T)= \left[A/\left[C+ \log\left(1+T_0/T\right)\right]+ B \vert p-p_c\vert\right]^{-1}.
\label{gamma}
\eeq
This form (with the parameters $A,\,B,\,C$, and $T_0$ being  adjusted by consistently fitting resistivity and specific 
heat data, see below), corresponds to the idea of a damping 
which increases by decreasing the temperature and is maximal at some critical doping $p_c$. The high-temperature limiting value 
is ruled by sub-leading temperature dependences of the fitting parameters. Since these are not constrained when fitting the 
low-temperature data, we do not address this issue in the present work. Eq.\,(\ref{gamma}) implies 
an anomalous-dissipative QCP, with a diverging $\gamma$ at $T=0$ and $p=p_c$. 
This latter assumption translates into the idea that the Planckian behavior may extend down to $T=0$, although this may not be 
experimentally assessed. As schematized in Fig.\,\ref{fig-0}(c), an increasingly larger $\gamma$ extends the linear resistivity to 
lower and lower temperatures.  By consistently fitting the resistivity and specific heat data (see below) we determine the
parameters  $T_0=200$\,K, $p_c=0.235$, $A=0.12$, $B=1.28$, $C=3.4$ for Nd-La$_{2-x}$Sr$_x$CuO$_4$,
and $T_0=120$\,K, $p_c=0.232$, $A=0.27$, $B= 5.97$, $C=1.1$ for Eu-La$_{2-x}$Sr$_x$CuO$_4$.  We 
find that the linear resistivity extends down to a few kelvins  (black curves and data in Fig. \ref{fig-1}). 

\begin{figure*}
\includegraphics[angle=0,scale=0.45]{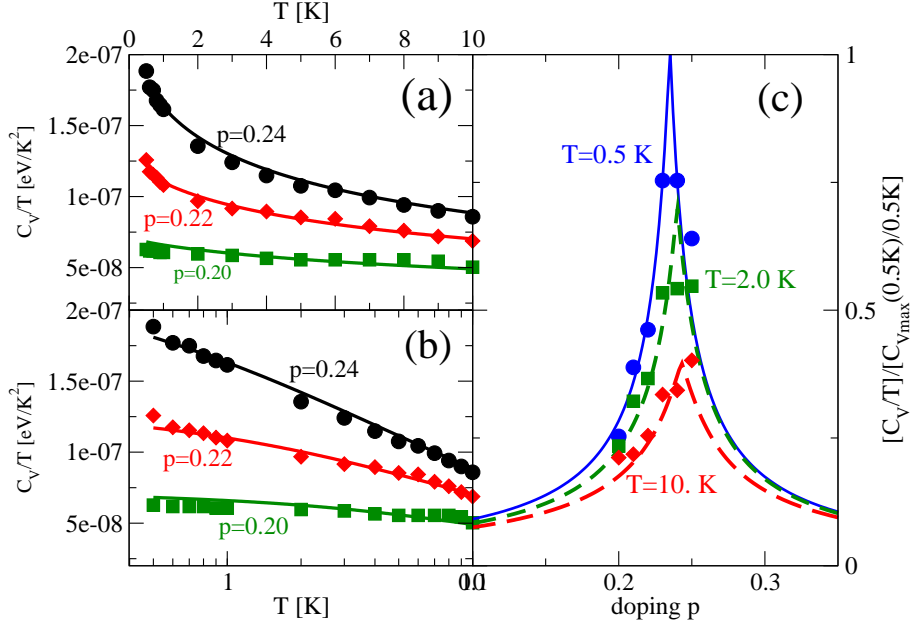} 
\caption{Temperature dependence of the low-temperature specific heat per unit cell (u.c.) over temperature  in Nd-La$_{2-x}$Sr$_x$CuO$_4$ samples with doping 
$p=0.24$ (black solid line and circles), $p=0.22$ (red solid line and diamonds), $p=0.20$ (green solid line and squares) 
in linear (a) and semilogarithmic (b) scales (the theoretical curves have been  rescaled by an overall factor 6.9, while keeping the relative weight 
at different doping and temperatures  fixed).
 (c) Doping dependence of  the low-temperature $C_V/T$  in 
Nd-La$_{2-x}$Sr$_x$CuO$_4$ samples at different temperatures $T=0.5,\, 2.0,\, 10.0$\,K. Both experimental and theoretical 
data are rescaled in order to have the maximum of the theoretical calculation at $T=0.5$ K normalised to one. The 
experimental data taken from Ref.\,\onlinecite{michon-2019} are represented by symbols, while the lines report our
theoretical calculations.}
\label{fig-2}
\end{figure*}

{\it --- Specific heat in cuprates ---}
The phenomenological assumption of a large $\gamma$, should  be validated by investigating its effect on other 
observables. In particular, since we claim that the main physical effect of a large damping is to shift the fluctuation 
spectral weight to lower energies, it is natural to expect a strong enhancement of the low-temperature specific heat. 
This is precisely what has been recently observed in other overdoped cuprates \cite{michon-2019}. Here we subtract from 
the observed specific heat the contribution of fermion quasiparticles. Despite the presence of a van Hove singularity, 
disorder, interplane coupling and electron-electron interactions smoothen this contribution. Thus fermion quasiparticles 
cannot account for the observed seemingly divergent specific heat. 

We argue instead that an enhancement of the boson contribution to the specific heat occurs if $\gamma$ 
obeys Eq.\,(\ref{gamma}). We start from the contribution of damped charge density fluctuations to free 
energy density $f_B=\frac{T}{V}\sum_{\Omega_\ell}\sum_{\qvec}\log\left[\mathcal{D}^{-1}(\qvec,\Omega_\ell)\right]$,
where $\mathcal{D}$ is the Matsubara propagator obtained after standard analytical continuation of Eq.\,(\ref{fluctuator}),
$\omega\to\mathrm i\Omega_\ell=2\pi\mathrm i\ell T$. 
Hence, we obtain the contribution of damped charge density fluctuations to the internal energy density $u_B$ and
to the specific heat 
\begin{equation}
\label{onion}
C_V^B=\frac{\partial u_B}{\partial{T}}=\frac{\partial}{\partial{T}}\left[\int_0^{\infty}\mathrm d\omega\,
\omega\,b(\omega)\,\rho_B(\omega)\right]
\end{equation}
where 
\begin{eqnarray*}
\rho_B(\omega)&=&
\frac{\gamma}{\pi^2{\bar{\nu}}}\log\left(\frac{1+W_+^2}{1+W_-^2}\right)+ \nonumber\\
&+&\frac{4\omega}{\pi^2{\bar{\nu}}{\gamma}{\overline{\Omega}}}
\left(\arctan W_+-\arctan W_-\right), \nonumber
\end{eqnarray*}
with $W_-=\frac{m_0}{\gamma\omega}-\frac{\omega}{\gamma\overline{\Omega}}$ and $W_+=W_- + \frac{\pi\bar{\nu}}{\gamma\omega}$,
plays the role of an effective spectral density. Fig.\,\ref{fig-2} shows that the 
enhancement of $\gamma(T,p)$ leading to a Planckian behavior in the low-$T$ resistivity, also induces a 
peak in the specific heat, due to the increase of low-energy boson degrees of freedom. Noticeably, the relative 
weight (height) of $C_V^B$ at the various temperatures is well captured by our approach. In particular, this feature is 
mostly ruled by the Bose distribution function in Eq.\,(\ref{onion}) and depends only little on the specific expression of 
$\gamma(T,p)$, provided enough spectral density is brought to frequencies $\omega\lesssim T$ with increasing $\gamma$. We 
also notice that the logarithmic temperature dependence of $\gamma$ mirrors in a nearly logarithmic behavior of $C_V/T$ 
[see Fig.\,\ref{fig-2} (a)].

{\it --- Discussion ---} The above analysis shows that two nontrivial features of the strange-metal behavior 
occurring near QCPs can be attributed to, and accounted for by, the damping parameter $\gamma$ only. We still lack a 
microscopic scheme to determine the doping and temperature dependence of $\gamma$, and, within the scope of 
the present work, we rely on the phenomenological expression of Eq.\,(\ref{gamma}). Therefore also the 
$T \log(T_0/T)$ behavior of the specific heat at $p_c$ is only phenomenologically captured by our theory. Nevertheless, 
we point out that our approach outlines a general paradigmatic change, shifting the relevance from the divergence of the 
correlation length $\xi$ to the increase (possibly divergence) of dissipation. This is precisely what renders our scheme 
different from the proposal of a local QCP put forward long ago in Ref.\,\onlinecite{qimiaosi}. In this latter case the 
critical behavior of the imaginary part (i.e., damping) of the self-energy of the critical fluctuations, is sublinear
$\mathrm i \gamma_0 \omega^{1-\alpha}$, which somehow rephrases our condition of an increasing damping at low energy 
scales by taking $ \mathrm i(\gamma_0/\omega)^\alpha \omega$  (i.e., $\gamma\sim \gamma_0/\omega^\alpha$), because of 
a diverging $\xi$. From our Eq. (\ref{gamma}) one can see that 
the assumption that at $p=p_c$ the scaling index in $T$ for $\gamma$ is zero, i.e., logarithmically divergent,
suggest that $\alpha\to 0$ and the challenge is to obtain this result 
without $\xi\to\infty$.

After momentum integration, a  similar frequency dependence characterises the singular 
dynamical interaction between quasiparticles mediated by the critical collective boson, in 
Ref.\,\onlinecite{chubukov-2020}, where a complete analysis of the complementary problem of the competition 
between pairing and non-FL metal at a QCP is reported. 

Of course, other, even more mundane, mechanisms might boost the increase of $\gamma$. 
In cuprates, for instance, $p_c$ occurs at or very near a van Hove singularity, which enhances the density 
of states of fermions, thereby increasing the Landau damping $\gamma$ of the fluctuations. Also the proximity to charge 
ordering might induce the reconstruction of the Fermi surface \cite{badoux}, thereby triggering an enhanced damping of 
the charge density fluctuations. All these are mechanisms worth being explored in the attempt to shape a microscopic 
theory for a damping-ruled violation of the FL behavior.

We thank Riccardo Arpaia, Lucio Braicovich, Claudio Castellani, and Giacomo Ghiringhelli for  stimulating discussions.
We acknowledge financial support from the University of Rome Sapienza, through the projects Ateneo 2017 (Grant No. 
RM11715C642E8370), Ateneo 2018 (Grant No. RM11816431DBA5AF), Ateneo 2019 (Grant No. RM11916B56802AFE), from the Italian 
Ministero dell'Universit\`a e della Ricerca, through the Project No. PRIN 2017Z8TS5B. G.S. acknowledges financial support
from the Deutsche Forschungsgemeinschaft under SE806/19-1


\begin{thebibliography}{99}
\bibitem{stewart} G. R. Stewart, Rev. Mod. Phys. {\bf 73}, 797---855 (2001).
\bibitem{Walmsley} P. Walmsley, {\it et al.}, Phys. Rev. Lett. 110, 257002 (2013).
\bibitem{anderson} P. W. Anderson, {\it The Theory of Superconductivity in the High Temperature Cuprates}, 1997, Princeton Legacy Library.
\bibitem{kastrinakis} G. Kastrinakis, Physica C {\bf 340} 119 (2000) 
\bibitem{CDG-ZP} C. Castellani, C. Di Castro, and M. Grilli, Z. Phys. B {\bf 103}, 137 (1996).
\bibitem{metzner-1990} C. Castellani, C. Di Castro, and W. Metzner,  Phys. Rev. Lett. {\bf 69}, 1703 (1992).
\bibitem{hlubina-1990} R. Hlublina and T.M. Rice, Phys. Rev. B {\bf 51}, 9253 (1995).
\bibitem{rosch-1990} A. Rosch, Phys. Rev. Lett. {\bf 82}, 4280 (1999).
\bibitem{belitz-kirkpatrick} M. Brando, D. Belitz, F. M. Grosche, and T. R. Kirkpatrick,  Rev. Mod. Phys. 
{\bf 88}, 025006 (2016).
\bibitem{varma} V. Aji and C. M. Varma, Phys. Rev. Lett. 99, 067003 (2007).
\bibitem{kivelson-scheurer} L. Dell'Anna and W. Metzner, Phys. Rev. Lett. {\bf 98}, 136402 (2007).
\bibitem{qimiaosi} Q. Si, S. Rabello, K. Ingersent, and J. Smith, Nature (London) {\bf 413}, 804 (2001).
\bibitem{coleman} P. Coleman, C. P\'epin, Q. Si, and R. Ramazashvili,  J. Phys.: Condens. Matter {\bf 13}, R723---R738 (2001).
\bibitem{burdin} S. Burdin, D. R. Grempel, and M. Grilli, Phys. Rev. B {\bf 75}, 224423 (2007).
\bibitem{sachdev-syk} A. A. Patel, J. McGreevy, D. P. Arovas, and S. Sachdev, Phys. Rev. X {\bf 8}, 021049 (2018).
\bibitem{legros-2019}  A. Legros, S. Benhabib,  W. Tabis, F. Lalibert\'e, M. Dion, M. Lizaire, B. Vignolle, D. Vignolles, H. 
Raffy, Z. Z. Li, P. Auban-Senzier, N. Doiron-Leyraud, P. Fournier, D. Colson, L. Taillefer, and C. Proust, Nat. Phys. 
{\bf 15}, 142 (2019) 
\bibitem{michon-2019} B. Michon, C. Girod, S. Badoux, J. Kamar\'\i k, Q. Ma, M. Dragomir, H. A.
Dabkowska, B. D. Gaulin, J.-S. Zhou, S. Pyon, T. Takayama, H. Takagi, S.
Verret, N. Doiron-Leyraud, C. Marcenat, L. Taillefer, T. Klein,  Nature {\bf 567}, 218222 (2019).
\bibitem{arpaia-2019} R. Arpaia, S. Caprara, R. Fumagalli, G. De Vecchi, Y. Y. Peng, E. Andersson, D.
Betto, G. M. De Luca, N. B. Brookes, F. Lombardi, M. Salluzzo, L. Braicovich,
C. Di Castro, M. Grilli, G. Ghiringhelli, Science {\bf 365}, 906-910 (2019).
\bibitem{seibold-2020}  G. Seibold, R. Arpaia, Y. Y. Peng, R. Fumagalli, L. Braicovich, C. Di Castro,
M. Grilli, G. Ghiringhelli, S. Caprara, Commun. Phys. to be published,
arXiv:1905.10232
\bibitem{CDG-1995} C. Castellani, C. Di Castro, and M. Grilli, Phys. Rev. Lett. {\bf 75}, 4650 (1995).
\bibitem{andergassen-2001} S. Andergassen, S. Caprara, C. Di Castro, and M. Grilli, Phys. Rev. Lett. {\bf 87}, 056401 (2001).
\bibitem{caprara-2002} S. Caprara, C. Di Castro, S. Fratini, and M. Grilli, Phys. Rev. Lett. {\bf 88}, 147001 (2002).
\bibitem{enss-2007} S. Caprara, M. Grilli, C. Di Castro, and T. Enss, Phys. Rev. B {\bf 75}, 140505(R) (2007).
\bibitem{miao-2020} H. Miao, G. Fabbris, C. S. Nelson, R. Acevedo-Esteves, Y. Li, G. D. Gu, T. Yilmaz, K. Kaznatcheev, 
E. Vescovo, M. Oda, K. Kurosawa, N. Momono, T. A. Assefa, I. K. Robinson, J. M. Tranquada, P. D. Johnson and M. P. M. 
Dean,  arXiv:2001.10294v2.
\bibitem{chubukov-2020} Artem Abanov and Andrey V. Chubukov, Phys. Rev. B {\bf 102}, 024524 (2020).
\bibitem{badoux} S. Badoux, W. Tabis, F. Lalibert\'e, G. Grissonnanche, B. Vignolle, D. Vignolles, J. B\'eard, D. 
A. Bonn, W. N. Hardy, R. Liang, N. Doiron-Leyraud, L. Taillefer, and C. Proust, Nature (London) {\bf 531}, 210 (2016).

\end{thebibliography}
\end{document}